\documentclass[12pt, prd, showpacs]{revtex4}
\usepackage{amssymb}
\usepackage{amsmath}
\usepackage{graphicx}

\input{tcilatex}

\begin{document}

\title{Acceleration of particles near the inner black hole horizon}
\author{O. B. Zaslavskii}
\affiliation{Department of Physics and Technology, Kharkov V.N. Karazin National
University, 4 Svoboda Square, Kharkov, 61077, Ukraine}
\email{zaslav@ukr.net }

\begin{abstract}
We study the possibility of obtaining unbound energy $E_{c.m.}$ in the
centre of mass frame when two particles collide near the inner black hole
horizon. We consider two different cases - when both particles move (i) in
the same direction or (ii) in the opposite ones. We also discuss two
different versions of the effect - whether an infinite energy can be
released in the collision (strong version) or the energy $E_{c.m.}$ is
finite but can be made as large as one likes (weak version). We demonstrate
that the strong version of the effect is impossible in both cases (i) and
(ii). In case (i) this is due to the fact that in the situation when $%
E_{c.m.}$ formally diverges on the horizon, one of particles passes through
the bifurcation point where two horizons meet while the second particle does
not, so collision does not occur. In case (ii), both particles hit different
branches of the horizon. The weak version is possible in both cases,
provided at least one of particles starts its motion inside the horizon
along the direction of spatial symmetry from infinity.
\end{abstract}

\maketitle




\section{Introduction}

In recent years, an interesting effect was discovered: when two particles
collide near the event horizon of a black hole, their energy $E_{c.m.}$ in
the centre of mass frame can grow unbound (so-called the BSW effect \cite%
{ban}). This provoked a series of papers in which properties of such
collisions were investigated in detail (see, e.g., the recent works \cite%
{spiral}, \cite{piat} and references therein). Meanwhile, for collisions
near the inner horizon the situation turned out to be contradictory. At
first, the possibility of the BSW effect on the inner horizons of the Kerr
and BTZ \ black holes was claimed in \cite{lake1}. Later, in a brief note 
\cite{lake2}, K. Lake claimed that although the formulas for the energy in 
\cite{lake1} are formally correct they are physically irrelevant since
actual collision between particles which would lead to infinite $E_{c.m.}$
cannot occur. A similar result was obtained in \cite{gp}, \cite{gpgc} for
the Kerr metric. However, in a recent work \cite{ch1}, the kinematics of
collisions inside black holes was discussed again with the conclusion that
the divergencies do occur on the inner horizon (provided the energy and
angular momentum of one particle are fine-tuned properly). The same
conclusion (but without consideration of the kinematics of collisions) is
made in \cite{ch2} for a cosmological horizon. Moreover, the results of \cite%
{ch2} would have implied that an infinite energy can be achieved during a
finite interval of time that looks unphysical.

The aim of the present paper is to give general explanation of the situation
with collisions near the inner black hole horizon valid for generic rotating
black holes and investigate the similar issue for charged nonrotating black
holes. (The same reasonings apply also to the nonextremal cosmological
horizons, so for definiteness we restrict ourselves by the inner black holes
ones). As was noticed in \cite{jl}, the counterpart of the BSW effect for
rotating black holes reveals itself for charged nonrotating ones. The latter
case captures main features of the phenomenon but consideration is simpler,
so at the beginning we discuss the motion of particles in the
Reissner-Nordstr\"{o}m metric. We show that the same conclusions apply to
any nonextremal spherically symmetric black holes having the inner horizon.

We also consider generic rotating nonextremal black holes (cf. \cite{prd}).
We show that the arguments of \cite{gp}, \cite{gpgc} were incomplete but the
result is correct in agreement with \cite{lake2}, so there is no BSW effect
with infinite energy on the inner or cosmological horizon in contrast to the
claims made in \cite{ch2}.

Apart from the BSW effect, we also consider another type of the effect
connected with the near-horizon collisions - the Piran and Shanam (PS) one 
\cite{ps}. The difference between both effects consists in that in the BSW
case both particles move in the same directions whereas in the PS case they
do it in opposite ones (see \cite{gen} for details).

In what follows, we need to distinguish two different versions of the
effects under discussion. I call it the "strong one" if the value of the
energy in the centre of mass frame is divergent in the point of collision.
And, it is called "weak version" if the energy is finite but can be made as
large as one likes. In the pioneering paper \cite{ban} where the BSW effect
was discovered, the energy in the centre of mass frame was found to be
infinite in the horizon limit, if special relationship between the energy
and momentum of one particle holds. However, later it was observed \cite%
{gpjl}, \cite{gpgc}, \cite{ted} that corresponding collision requires an
infinite proper time, so physically the infinite energy cannot be realized.
This observation was extended to the generic case in \cite{prd}. Apart from
this, it is pointed out in \cite{gpjl}, \cite{prd}, \cite{gpgc} that the BSW
effect is valid for nonextremal horizons, provided multiple scattering
occurs. In each collision the energy $E_{c.m.}$is finite but it can be made
as large as one likes if the radial coordinate of collision becomes closer
and closer to the horizon radius. (Additionally, it requires some special
relationship between the energy and angular momentum or the energy and
electric charge for one of colliding particles.) Thus for both types of the
event horizons (extremal and nonextremal) only the weak version of the BSW
effect can be realized.

In this terminology, when it applies to inner horizons, refutations made in 
\cite{lake2} and \cite{gp}, \cite{gpgc} concern the strong version of the
BSW effect only but they say nothing about the possibility of the weak
version. This will be done below. The claim of \cite{ch1} can be
reformulated by saying that the reason why the strong version of the BSW
effect is not realized is due to an infinite time required for the critical
particle to reach the horizon. We will see that this is incorrect since the
corresponding time is finite and the reason why the strong version of the
BSW effect does not happen is quite different. The result of the paper \cite%
{ch2} for the cosmological horizon is even more unphysical since it would
have meant that the strong version of the effect of collision does occur.

It is worth noting that there are possible limitations on the BSW effect due
to backreaction and gravitational radiation near the event horizon \cite{ted}%
, \cite{vit}. Similar factors can come into play near the inner horizon as
well. At present, their role is not fully understood and we refrain from
discussing this important physical issue that needs separate investigation.

\section{Motion inside the Reissner-Nordstr\"{o}m black hole}

At first, let us consider the metric of the Reissner-Nordstr\"{o}m black hole%
\begin{equation}
ds^{2}=-dt^{2}f+\frac{dr^{2}}{f}+r^{2}d\omega ^{2}.  \label{met}
\end{equation}%
Here $d\omega ^{2}=\sin ^{2}\theta d\phi ^{2}+d\theta ^{2}$, $f=1-\frac{2M}{r%
}+\frac{Q^{2}}{r^{2}}$, $M$ is the black hole mass, $Q$ is its charge, $M>Q$%
. We use the geometrical system of units with $c=G=1$ ($G$ is the
gravitational constant, $c$ is the speed of light). The function $f(r)=0$ at 
$r_{-}=M-\sqrt{M^{2}-Q^{2}}$ (the inner horizon) and $r_{+}=M+\sqrt{%
M^{2}-Q^{2}\text{ }}$ (the event horizon).

We are interested in the region $r_{\_}\leq r<r_{+}\,\ $\ where $f=-g\leq 0$%
. In this region, the coordinate $r\equiv T$ has timelike character and $%
t\equiv y$ is spacelike. Then, the metric takes the form 
\begin{equation}
ds^{2}=-\frac{dT^{2}}{g(T)}+g(T)dy^{2}+T^{2}d\omega ^{2}\text{.}  \label{m}
\end{equation}

Let a particle have the charge $q$ and the rest mass $m$. For simplicity, we
consider its motion along the $y$ -direction. As the metric does not depend
on $y$, the momentum $P_{y}\equiv P$ is conserved. We assume that $r=T$
decreases, so the region under discussion is the $T_{-}$ region in the
Novikov's terminology \cite{n}. The equations of motion read 
\begin{equation}
m\dot{y}=\frac{X}{g}\text{, }X=(P+\frac{qQ}{T}),  \label{x}
\end{equation}%
\begin{equation}
\dot{T}=-\sqrt{g+\frac{X^{2}}{m^{2}}}  \label{T}
\end{equation}%
where dot denotes differentiation with respect to the proper time $\tau $.
It follows from (\ref{T}) that%
\begin{equation}
\tau =m\int_{r}^{r_{1}}\frac{dr}{\sqrt{m^{2}g+X^{2}}}  \label{tau}
\end{equation}%
where we an initial value moment of time $r_{1}$such that $r_{-}<r\leq
r_{1}<r_{+}$.

The choice of signs in (\ref{x}) takes into account that $P=-E$ where $E$
has the meaning of the conserved energy in $R$ region. It follows from (\ref%
{x}), (\ref{T}) that 
\begin{equation}
\frac{dy}{dT}=-\frac{1}{g}\frac{X}{\sqrt{m^{2}g+X^{2}}}\text{.}  \label{yT}
\end{equation}

It follows from (\ref{yT}) that 
\begin{equation}
y=\int_{r}^{r_{1}}\frac{1}{g}\frac{Xdr}{\sqrt{m^{2}g+X^{2}}}+y(r_{1}).
\label{y1}
\end{equation}

If there are two particles with charges $q_{1}$ and $q_{2}$ and masses $m$,
the energy in the centre of mass frame is equal to (see \cite{jl}, \cite{prd}
and references therein)

\begin{equation}
\frac{E_{cm}^{2}}{2m^{2}}=1+\frac{Z_{1}Z_{2}-X_{1}X_{2}}{gm^{2}}.  \label{e}
\end{equation}%
Here, $i=1,2$ enumerates particles, 
\begin{equation}
Z_{i}=\sqrt{X_{i}^{2}+m^{2}g}\text{,}  \label{xz}
\end{equation}%
$X_{1}X_{2}>0$ if particles move in the same direction and $X_{1}X_{2}<0$ if
the particles move in the opposite ones. If the charges $q_{1}$ and $q_{2}$
have the same sign, Coloumb repulsion somewhat complicates the picture of
collision. To avoid such unnecessary details, we assume that, say, $q_{2}=0$%
. The potential divergencies can occur in the limit $g\rightarrow 0$ only,
i.e. near the inner horizon where $r\rightarrow r_{-}$.

Now, we examine the possibility of two effects separately.

\section{BSW effect ($X_{1}X_{2}>0$).}

\subsection{Energy of collision}

Here, one should distinguish two types of particles. By analogy with
previously used terminology \cite{prd}, we call a particle usual if $%
X_{i}(r_{-})\neq 0$ and critical if $X_{i}(r_{-})=0$ ($i=1,2$). In the
latter case 
\begin{equation}
X(r)=-\frac{q_{1}Q}{rr_{-}}(r-r_{-})\text{.}  \label{xc}
\end{equation}

If both particles are usual, near the horizon $Z_{i}\approx \left\vert
X_{i}\right\vert +\frac{m^{2}g}{2\left\vert X_{i}\right\vert }$ and,
according to (\ref{e}), $E_{cm}^{2}$ remains finite, the effect is absent.
If both particles are critical, near the horizon $X_{i}\sim r-r_{-}\sim g$, $%
Z_{i}\approx m\sqrt{g}$, and the effect is also absent. The only potential
case of interest arises when the particles belong to different types. Say,
particle 1 is critical and particle 2 is usual. Then,

\begin{equation}
\frac{E_{cm}^{2}}{2m^{2}}\approx \left\vert X_{2}(r_{-})\right\vert \frac{1}{%
\sqrt{g}m}\rightarrow \infty  \label{ee}
\end{equation}%
when $r\rightarrow r_{-}$. However, the crucial point consists in that one
should check whether a collision as such occurs near the inner horizon.

\subsection{Trajectories in original coordinates}

For what follows, we need explicit asymptotic behavior of space-time
trajectories near the horizon. For a usual particle, one can easily obtain
from (\ref{x}) - (\ref{xz}) that

\begin{equation}
t=y\approx C-\frac{signX(r_{-})}{2\kappa _{-}}\ln (r-r_{-})\text{,}
\label{tu}
\end{equation}%
\begin{equation}
r=T\approx r_{-}+\frac{\left\vert X(r_{-})\right\vert }{m}(\tau _{-}-\tau ).
\label{ru}
\end{equation}%
Here, $\kappa _{-}=\frac{1}{2}\left( \frac{dg}{dr}\right) _{\mid r=r_{-}}$
has a meaning of the surface gravity of the inner horizon, $C$ and $\tau
_{-} $ are constants.

For the critical particle,%
\begin{equation}
y-y_{-}\approx A\sqrt{r-r_{-}}\text{,}  \label{yr}
\end{equation}%
\begin{equation}
\tau -\tau _{-}\approx -B\sqrt{r-r_{-}}\text{,}  \label{tr}
\end{equation}%
where $y_{-}$, $A$, $B$ are constants,%
\begin{equation}
A=\frac{-qQ}{\sqrt{2}m\kappa _{-}^{3/2}r_{-}^{2}},
\end{equation}%
\begin{equation}
B=\frac{\sqrt{2}}{\kappa _{-}^{1/2}}\text{.}
\end{equation}%
It is seen from (\ref{tu}), (\ref{ru}) that the proper time is finite both
for usual and critical particles (in contrast to the results described in
the end of Sec. III of Ref. \cite{ch1}). For usual particles, $\frac{dt}{%
d\tau }\rightarrow \infty $ or $\frac{dt}{d\tau }\rightarrow -\infty $,
depending on the sign of the momentum $X$ in full analogy with the situation
for the Kerr metric \cite{gp}, \cite{gpgc}. Meanwhile, for critical
particles, $\frac{dt}{d\tau }$ remains finite. It was concluded in \cite{gp}%
, \cite{gpgc} that the collision between the critical particle and a usual
one (which is necessary for divergences of $E_{cm}^{2}$) cannot occur since
the difference in the variable $t$ is infinite for them. Such a conclusion
looks plausible but not quite rigorous since the coordinate system used in
the analysis becomes degenerate near the horizon, so the behavior of
coordinates gives essentially incomplete information about the process.
Apart from this, we want to examine not only the strong version of the BSW
effect but also the weak one. To give full self-consistent picture, one
should reformulate the metric in well-behaved coordinates.

\subsection{Trajectories in Kruskal coordinates}

In original coordinates (\ref{met}), the metric coefficients become
ill-behaved near the horizon. To remedy this drawback, one is led to using
coordinates in which the metric coefficients are analytical near the
horizon. We take advantage of the Kruskal-like coordinates which were
introduced in \cite{kr} for the Schwarzschild metric. Now, we apply
corresponding formulas to the region inside the horizon where the metric
reads

\begin{equation}
ds^{2}=-FdUdV+r^{2}d\omega ^{2}\text{,}
\end{equation}%
and%
\begin{equation}
U=\exp [-\kappa _{-}(t-r^{\ast })]\text{,}  \label{u}
\end{equation}%
\begin{equation}
V=\exp [\kappa _{-}(t+r^{\ast })]\text{,}  \label{v}
\end{equation}%
the tortoise coordinate%
\begin{equation}
r^{\ast }=\int \frac{dr}{g}\text{.}  \label{r}
\end{equation}%
\begin{equation}
F=g\kappa _{-}^{2}\exp (-2\kappa _{-}r^{\ast })\text{.}  \label{F}
\end{equation}%
Near the horizon, the tortoise coordinate diverges, 
\begin{equation}
r^{\ast }\approx \frac{1}{2\kappa _{-}}\ln (r-r_{-})+r_{0}^{\ast }
\label{tort}
\end{equation}%
where $r_{0}^{\ast }$ is a constant.

The coordinates $U$ and $V$ take finite values near the horizon. The surface 
$U=0$ corresponds, say, to the left inner horizon in the standard
Carter-Penrose diagram while $V=0$ corresponds to the right one. Near any of
two horizons, 
\begin{equation}
g\approx 2\kappa (r-r_{-}),  \label{gk}
\end{equation}%
$F$ is finite. As a result, the zeros in the numerator and denominator in (%
\ref{F}) compensate each other, so the metric coefficient $F$ is finite and
nonzero on the horizon.

Then, it follows from formulas (\ref{u}) - (\ref{tort}) that 
\begin{equation}
\frac{U}{V}=\exp (-2\kappa _{-}t)
\end{equation}%
and, near the horizon%
\begin{equation}
UV\approx const(r-r_{-})\text{.}
\end{equation}

For usual particles near, say, the left horizon the value of coordinate $V$
is finite$,V\neq 0$, $t\rightarrow \infty $, $r-r_{-}\sim \exp (-2\kappa
t)\rightarrow 0$, $U\rightarrow 0$. For critical particles, it is seen from (%
\ref{yr}) that $t$ is finite, $V\sim U\sim \sqrt{t_{-}-t}\sim \sqrt{\tau
_{-}-\tau }\rightarrow 0$. Thus, critical and usual particles have at the
horizon different values of $V$ and, therefore, cannot collide there. This
confirms the statements of \cite{lake2} and \cite{gp}, \cite{gpgc}.

Here, we comment shortly on the corresponding claims made in \cite{ch1}.
These authors rely on the properties of the critical particle and claim that
it only asymptotically spirals onto the horizon for an infinite proper time
similarly to the situation analyzed in \cite{ted}, \cite{gp}. Then,
according to Ref. \cite{ch1}, collision with an infinite energy would occur
at an arbitrary point of the inner horizon, and only an infinite proper time
prevents it from being actual event. Meanwhile, the corresponding
observations in Refs. \cite{ted}, \cite{gp} refer to the situation in $R-$%
region outside the extremal event horizon. They do not apply to particles'
motion near the inner horizon which is nonextremal. Mathematically, the
integral (\ref{tau}) converges even in the critical case since the function $%
g$ has the simple zero. Therefore, the proper time for the critical particle
to reach the horizon is finite in contrast to the claim made in Ref. \cite%
{ch1}. Moreover, for the Reissner-Nordstr\"{o}m (or Reissner-Nordstr\"{o}m -
de Sitter one like in Ref. \cite{ch2}) the effect reveals itself even for a
zero angular momentum, so there is no any spiraling at all. Actually, the
mechanism preventing the strong version of the effect is completely
different and this will be shown below.

\subsection{Critical particle and bifurcation point}

Now, it is worth noting that motion of the critical particle admits simple
geometrical interpretation. It follows from (\ref{T}) that the critical
particle reaches the horizon in finite proper time (in contrast to the
situation with the event horizon \cite{gp}, \cite{jl}, \cite{prd}, \cite{ted}%
). It means that a particle should cross the horizon. However, in the $R$
region, where $f>0$, such a particle cannot be situated in the immediate
vicinity of the horizon. Indeed, in that region, eq. (\ref{T}) turns to%
\begin{equation}
\left( \dot{r}\right) ^{2}=\frac{X^{2}}{m^{2}}-f\text{.}  \label{rr}
\end{equation}

\bigskip Near the horizon, for $r<r_{-}$, $f\sim r_{-}-r$. For the critical
particle, $X\sim r_{-}-r$. Thus, near $r_{-}$ for $r<r_{-}$, the right hand
side of (\ref{rr}) becomes negative and motion is impossible. The same
conclusion is valid for rotating nonextremal black holes \cite{prd}, \cite%
{ch1}. Thus, the critical particle cannot find itself in the $R$ region. The
only possibility that remains for it is to enter $T$ region again. But for
the Reissner-Nordstr\"{o}m nonextremal metric the only possible path for it
passes through the bifurcation point where two horizons meet. Meanwhile, a
usual particle reaches the horizon somewhere in an intermediate point, so
these points are separated geometrically. The corresponding situation is
represented at the part of the Carter-Penrose diagram in Fig. 1 where B is
the bifurcation point.


It is instructive to remind that if both particles collide approaching the
extremal horizon from the $R$ region (outside the event horizon) \cite{ban},
the critical particle plays a crucial role in the BSW effect but in such a
case there is no bifurcation point at all.

\subsection{Critical particle and speed of motion}

In addition to geometrical properties of the trajectory of the critical
particle near the horizon, it is instructive to look at the kinematic ones.
One can define the velocity with respect to an observer who remains at rest: 
$v=\frac{dl}{d\tau }$, $dl=dy\sqrt{g}$, $d\tau =\frac{dT}{\sqrt{g}}$, so $v=g%
\frac{dy}{dT}$. Then, after simple manipulations, one obtains from (\ref{x}%
), (\ref{T}) that 
\begin{equation}
\frac{X}{m}=\pm \sqrt{g}\frac{v}{\sqrt{1-v^{2}}}.
\end{equation}

For usual particles, $X(r_{-})\neq 0$, so near the horizon $v\approx
1-\kappa _{-}\frac{m^{2}}{X^{2}(r_{-})}(r-r_{-})$ where we took into account
(\ref{gk}). For the critical particle, $X(r)\sim r-r_{-}$ near the horizon,
so $v\sim \frac{X}{\sqrt{g}}\sim \sqrt{r-r_{-}}\rightarrow 0$. Thus, the
critical particle not only has the velocity different from that of light -
moreover, it approaches the horizon with almost vanishing velocity.

For particles which collide near the horizon from the outside, the BSW
effect received a simple explanation based on kinematic properties \cite{k}.
Namely, in the static frame a usual particle have the velocity $v\rightarrow
1$ near the horizon, whereas for critical ones $v\neq 1$ near the horizon.
Then, the relative velocity tends to the speed of light, the Lorentz factor
grows unbound and the energy in the comoving frame tends to infinity.
Roughly speaking, a quick particle overtakes a slow one that results in the
almost infinite energy of collision in the centre of mass frame. However,
collision with the literally infinite energy in the centre of the type
represented in Fig. 1 is impossible.

\section{Are near-horizon collisions with finite but unbound energy possible?%
}

Thus we saw that the collisions with the infinite energies cannot occur. In
other words, the strong version of the BSW effect cannot be realized
physically. Meanwhile, one can ask, whether we can at least arrange
collisions not exactly on the horizon but somewhere in its vicinity with the
energy which would grow while approaching the horizon (the weak version of
the BSW effect). It is instructive to remind that for collisions which occur
from the outside the horizon, such situation is indeed possible not only for
extremal black holes but also for nonextremal ones \cite{gpjl}, \cite{gpgc}, 
\cite{prd}.

Let particle 1 be the critical one, as before. For a generic particle 2 that
would hit the horizon with generic value of coordinate $V$ both particles
are still separated in agreement with Fig. 1. Meanwhile, we can arrange
collision not exactly on the horizon but somewhere in its vicinity. We can
assume that an usual particle which without collision would hit the right
horizon at point A with $V\rightarrow 0$, now collides with particle 1 at
point C - see Fig. 2.


It means that collision is adjusted in such a way that points A. B and C are
close to each other and the value of $V$ in point $C$ is small.
Correspondingly, the value $r_{0}$ in the point of collision is close to $%
r_{-}$. Then, according to eq. (\ref{ee}), the energy becomes arbitrarily
large.

Meanwhile, there is a kinematic issue to be clarified. The situation under
discussion implies that particle 2 possesses two important properties: (i)
it is usual, so $X_{2}(r_{-})\neq 0,$(ii) in the absence of collision it
would hit the horizon with arbitrarily small $V$. Are properties (i) and
(ii) mutually consistent? Now we will show that the answer is "yes".

It follows from eqs. (\ref{tu}) (with $X_{2}>0$ for definiteness), (\ref{v})
and (\ref{tort}) that near the horizon $V_{2}\approx e^{\kappa _{-}C}$where
for simplicity we put $r_{0}^{\ast }=0$ for the constant in (\ref{tort}).
Therefore, for generic finite $C$ particle 2 cannot have small $V$ near the
horizon. However, this becomes possible if the constant $C$ is chosen to be
(in modulus) large and negative that implies that the constant $y(r_{1})$ in
(\ref{y1}) is also large and negative. A typical trajectory of such a kind
is given in Fig. 2. Particle 2 passes nearly to the left corner of the
Carter-Penrose diagram, keeps moving closely to the left horizon and would
hit the right horizon in point A in the absence of collision.

Taking into account eqs. (\ref{tu}) - (\ref{ru}) and (\ref{u}), (\ref{v}),
one can write for coordinates of both particles (critical and usual) near
the horizon%
\begin{equation}
U_{1}\approx U_{10}\sqrt{r_{0}-r_{-}}\text{, }V_{1}\approx V_{10}\sqrt{%
r_{0}-r_{-}}\text{, }  \label{u1}
\end{equation}%
\begin{equation}
U_{2}\approx U_{20}(r_{0}-r_{-})\text{, }V_{2}\approx V_{20}  \label{u2}
\end{equation}%
where 
\begin{equation}
U_{10}=e^{-\kappa _{-}y_{-}}\text{, }V_{10}=e^{\kappa y_{-}}\text{, }%
U_{20}=e^{^{-\kappa _{-}C}}\text{, }V_{20}=e^{\kappa _{-}C}\text{.}
\label{uv}
\end{equation}

It is seen that the condition of collision $U_{1}=U_{2}$, $V_{1}=V_{2}$ has
one solution for which 
\begin{equation}
U_{10}=U_{20}\sqrt{r_{0}-r_{-}},V_{20}=V_{10}\sqrt{r_{0}-r_{-}}\text{, }%
e^{\kappa (C-y_{-})}=\sqrt{r_{0}-r_{-}}  \label{=}
\end{equation}%
where $r_{0}$ is a point of collision and eq. (\ref{yr}) was taken into
account. This is achieved at the expense of large and negative $C$. It is
obvious that one can deform slightly the mutual disposition of particles in
Fig. 2 in such a way that particle 1 ceases to be critical but remains
near-critical that does not change our main conclusions.

Then, the collision occurs near the bifurcation point with small (although
nonzero) coordinates $U$ and $V$ and with the large (although finite) energy 
$E_{cm}$. Thus the BSW\ effect is present in the weak version under
discussion. However, this imposes conditions not only on particle 1 which
must be critical or near-critical but requires also that an usual particle
belong to a special class of trajectories. More precisely, in eq. (\ref{y1}%
), (\ref{tu}) constants $y(r_{1})$ and $C$ should be large and negative.
This means that a usual particle 2 starts its motion from "almost" infinity
in the sense that $\left\vert y\right\vert $ is large. It starts from large
and negative $y$ and moves from the left to the right if $X_{2}>0$ (so $\dot{%
y}>0$) and vice versa.

It is instructive to compare the situation of infinitely growing energies
for collisions on the inner end event horizons. In both cases, one of
particle should be critical or near-critical. In doing so, the velocity of a
usual particle with respect to $(r,t)$ coordinates approaches the speed of
light whereas the velocity of the critical one does not. Meanwhile, there is
difference in geometric properties between both cases. Near the event
horizon, the four-velocity of a critical particle has the component along
one horizon generator much larger than along the other one \cite{gen}. For
the inner horizon the situation is opposite as (without mathematical rigor)
can be seen from Fig. 2 clearly.

\section{PS effect ($X_{1}X_{2}<0)$}

Now, let us discuss the case when both colliding particle move in the
opposite directions (the PS effect). We would also like to emphasize the
difference between the PS\ effect on the event horizon \cite{ps} and on the
inner one. In the first case, one has to combine the metric of a black hole
with the state of a particle moving away from the horizon instead of a usual
picture of a particle approaching the horizon in the course of gravitational
collapse. This requires the choice of very special initial conditions - say,
such a particle should acquire its momentum (or be created ) in some other
precedent process. By contrary, inside the event horizon motion in both
directions are physically equivalent, so in this region the PS effect is
more natural than outside.

Hereafter, we assume that $X_{1}>0$ (so $\dot{y}>0)$ and $X_{2}<0$ (so $\dot{%
y}<0$). In contrast to the collisions in $R$ region, now particles move
along the legs of a cylinder and for both of them $\dot{r}<0$. It is seen
from (\ref{e}) that if both particles are usual, $E_{cm}^{2}\sim
g^{-1}\rightarrow \infty $ near the inner horizon. If both particles are
critical, $X_{i}\sim g$, $Z_{i}\sim \sqrt{g}$, and $E_{cm}^{2}$ is finite,
so this case is not interesting. If, say, particle 1 is critical and
particle 2 is usual, $X_{1}\sim g$, $X_{2}(r_{-})\neq 0$, $Z_{1}\sim \sqrt{g}
$, $Z_{2}(r_{-})\neq 0$, $E_{cm}^{2}\sim g^{-1/2}\rightarrow \infty $, so
the energy diverges, although more slowly than in the case when both
particles are usual. Thus, it is interesting that, in contrast to the BSW
effect, the most rapid grow of the energy $E_{c.m.}$ occurs when none of
particles is critical.

Now, we must analyze the possibility of collision from the kinematic
viewpoint.

\subsection{Impossibility of strong version}

First of all, we must check that the strong version of the effect is
impossible. Let us consider different cases separately.

\subsubsection{Particle 1 is critical, particle 2 is usual.}

Then, the situation is presented at Fig. 3 which is similar to Fig. 1, so
collision with the infinite energy does not occur.


\subsubsection{Both particles are usual.}

Using the asymptotic form of trajectories (\ref{tu}), (\ref{ru}) and eqs. (%
\ref{u}), (\ref{v}) we find that%
\begin{equation}
U_{1}\approx U_{10}(r-r_{-})\text{, }V_{1}\approx V_{10}\text{,}  \label{u1p}
\end{equation}%
\begin{equation}
U_{10}=e^{-\kappa _{-}C_{1}}\text{, }V_{10}=e^{\kappa _{-}C_{1}}.
\end{equation}

For particle 2, 
\begin{equation}
U_{2}\approx U_{20}\text{, }V_{2}\approx V_{20}(r-r_{-})\text{.}
\end{equation}%
\begin{equation}
U_{20}=e^{-\kappa _{-}C_{2}}\text{, }V_{20}=e^{\kappa _{-}C_{2}}.  \label{=p}
\end{equation}

If constants $C_{1}$, $C_{2}$ (hence, also $U_{10}$, $U_{20}$, $V_{10}$, $%
V_{20}$) are all finite, it is seen that in the horizon limit when $%
r\rightarrow r_{-}$, $U_{1}\rightarrow 0$, $V_{1}\neq 0$ and $%
V_{2}\rightarrow 0$, $U_{1}\neq 0$. This means that particle 1 hits the left
horizon whereas particle 2 hits the right one, so again collision with the
infinite energy does not happen. The situation is represented at Fig. 4.


\subsection{Weak version}

Obviously, if the particles start having some separation and move along the $%
y$ axis to meet each other, head-on collision is inevitable and occurs in
some finite proper time. In doing so, the energy is finite. However, it will
be seen now that the energy can be made as large as one wants if one makes
the point of collision closer and closer to the horizon. Again, we analyze
different cases separately.

\subsubsection{Particle 1 is critical, particle 2 is usual.}

Then, eqs. (\ref{u1}) - (\ref{=}) are valid and the corresponding analysis
applies. The situation is represented at Fig. 5 which is similar to Fig. 2.


\subsubsection{Both particles are usual.}

Then, near the horizon eqs. (\ref{u1p}) - (\ref{=p}) are valid. One can
arrange collision ($U_{1}=U_{2}$, $V_{1}=V_{2}$) if one chooses $%
U_{20}=U_{10}(r_{0}-r_{-})$, $V_{10}$.$=V_{20}(r_{0}-r_{-})$ or,
equivalently, $e^{\kappa _{-}C_{1}}=e^{\kappa _{-}C_{2}}(r_{0}-r)$ where $%
r_{0}$ corresponds to the point of collision. This can be also obtained in
original coordinates (\ref{m}) directly, taking into account the asymptotic
form of trajectories from (\ref{x}), (\ref{T}). We can rewrite the relation
between the constants as

\begin{equation}
C_{2}\approx C_{1}-\frac{1}{\kappa _{-}}\ln (r_{0}-r_{-}).  \label{c2}
\end{equation}%
For any $r_{0}\neq r_{-}$, the energy $E_{cm}$ is finite. However, choosing $%
r_{0}-r_{-}<<r_{-}$ and simultaneously taking trajectories with larger and
larger $C_{2}$, we can indeed achieve collision with $E_{c.m.}$ as large as
we want.

The situation is represented at Fig. 6 (collision near the generic point of
the horizon, $C_{1}\rightarrow -\infty $, $C_{2}$ is finite)


and Fig. 7 (collision near the bifurcation point, $C_{1}\rightarrow -\infty $%
, $C_{2}\rightarrow \infty $). It is worth reminding that large and negative
(positive) $C$ corresponds to particles which start from the left (right)
infinity in terms of the coordinate $y$.


\section{Intermediate case ($X_{1}X_{2}=0$)}

Now, we will discuss the case intermediate between BSW and PS effects -
namely, when one of $X_{i}$ vanishes (say, $X_{1}=0$). This means that $%
y_{1}=const$, the particle remains at rest with respect to this coordinate
system. (Meanwhile, the geometry evolves since the region is nonstationary, $%
\dot{r}<0$.) Then, collision between particles can be thought of as a
counterpart of that in the Kerr background when one particle is on the
circular orbit \cite{kerr}.

It follows from (\ref{x}) that the condition $X_{1}=0$ for all $r$ is
possible only for a special case $P_{1}=0=q_{1}$. Actually, particle 1 is
formally critical since $X_{1}(r_{-})=0$ (although now this holds not only
for $r_{-}$ but for all $r$). Therefore, the above analysis applies. In the
present case, $Z_{1}=m\sqrt{g}$, 
\begin{equation}
\frac{E_{cm}^{2}}{2m^{2}}=1+\frac{Z_{2}}{m\sqrt{g}}=1+\sqrt{1+\frac{X_{2}^{2}%
}{m^{2}g}}\text{.}
\end{equation}

If particle 2 is critical, $E_{cm}^{2}$ remains finite in accordance with
what is said above. If particle 2 is usual, $X_{2}(r_{-})\neq 0$, $\frac{%
E_{cm}^{2}}{2m^{2}}\rightarrow \infty $ when the moment of collision $%
r_{0}\rightarrow r_{-}$. For any finite $y_{1}$, collision occurs outside $%
r_{-}$. However, taking $r_{0}-r_{-}\rightarrow 0$ and, correspondingly, $%
\left\vert y_{1}\right\vert \rightarrow \infty $, one can achieve the
collision near the horizon with $E_{cm}^{2}$ as large as one wishes, so
again we have the weak version of the effect.

\section{Generalization, extension to generic rotating black holes}

We can consider more general spherically symmetric black hole metrics of the
form

\begin{equation}
ds^{2}=-dt^{2}f_{1}+\frac{dr^{2}}{f_{2}}+r^{2}d\omega ^{2}\text{.}
\end{equation}

Let us suppose that such a black hole is nonextremal and near the inner
horizon $f_{1}\sim f_{2}\sim r_{-}-r$ . Then, all above consideration
applies since it was based on the asymptotic properties of the metric
coefficients, while the explicit form of the Reissner-Nordstr\"{o}m metric
was irrelevant. Repeating all reasonings step by step we arrive at the
conclusion that the strong version of the BSW and PS effects is forbidden
whereas the weak one is allowed.

The situation with rotating black holes is somewhat more complicated. The
metric reads%
\begin{equation}
ds^{2}=-N^{2}dt^{2}+g_{\phi \phi }(d\phi -\omega dt)^{2}+dl^{2}+Bdz^{2}\text{%
.}  \label{z}
\end{equation}

We assume that all metric coefficients do not depend on $t$ and $\phi $. We
also assume that $\theta =\frac{\pi }{2}$ ($z=0$) is the symmetry plane and
restrict ourselves by motion in this plane. Then, the equations of motion
read \cite{prd}%
\begin{equation}
\dot{t}=\frac{E-\omega L}{N^{2}},  \label{trot}
\end{equation}%
\begin{equation}
\dot{\phi}=\frac{L}{g_{\phi \phi }}+\frac{\omega (E-\omega L)}{N^{2}},
\label{frot}
\end{equation}%
\begin{equation}
\dot{l}^{2}=\frac{(E-\omega L)^{2}}{N^{2}}-1-\frac{L^{2}}{g_{\phi \phi }}
\label{lrot}
\end{equation}%
where $L$ is the angular momentum, $E$ is the energy and the value $\theta =%
\frac{\pi }{2}$ is put in all metric coefficients. Inside the horizon we
must replace $N^{2}$ by $-g<0$ and the proper distance by the proper time $%
\tau $.\ Now, the metric takes the form%
\begin{equation}
ds^{2}=-d\tau ^{2}+gdy^{2}+g_{\phi \phi }(d\phi -\omega dy)^{2}+Bdz^{2}\text{%
.}
\end{equation}%
where $t=y$ is a spatial coordinate.\ 

Then, near the inner horizon,%
\begin{equation}
\sqrt{g}\approx \kappa _{-}\tau +O(\tau ^{3})\text{,}  \label{gz}
\end{equation}%
\begin{equation}
\omega \approx \omega _{i}+A(z)\tau ^{2}.  \label{w}
\end{equation}%
where $\kappa _{-}$ and $\omega _{i}$ are constants. The quantity $\kappa
_{-}$ has the meaning of the surface gravity of the inner horizon.
Derivation of (\ref{gz}), (\ref{w}) can be found in \cite{vis} with obvious
replacement $l\rightarrow \tau .$

It is also convenient to introduce the coordinates $x=\frac{1}{4}\tau ^{2}$
and $\tilde{\phi}=\phi -\omega _{i}y$, $y=\frac{\tilde{y}}{2\kappa _{-}}.$
Then,%
\begin{equation}
ds^{2}\approx -\frac{dx^{2}}{x}+d\tilde{y}^{2}x+g_{\phi \phi }^{-}(d\tilde{%
\phi}-\tilde{A}xd\tilde{y})^{2}+B^{-}dz^{2}\text{,}
\end{equation}%
\begin{equation}
g_{\phi \phi }^{-}=g_{\phi \phi \mid x=0}\text{, }B^{-}=B_{\mid x=0}\text{, }%
\tilde{A}=\frac{2A_{\mid x=0}}{\kappa _{-}}\text{.}
\end{equation}%
Near the horizon, one can obtain that the asymptotic behavior of space-time
trajectories (\ref{tu}) - (\ref{tr}) is still valid.

Further, let us introduce%
\begin{equation}
x^{\ast }=\ln x\text{, }u=x^{\ast }-\tilde{y}\text{, }v=\tilde{y}+x^{\ast }%
\text{,}
\end{equation}%
\begin{equation}
U=\exp (\frac{u}{2})\text{, }V=\exp (\frac{v}{2})\text{.}
\end{equation}%
Then, it is seen that the metric becomes analytical near the horizon where%
\begin{equation}
ds^{2}\approx -4dUdV+g_{\phi \phi }^{-}[d\tilde{\phi}-\tilde{A}%
(UdV-VdU)]^{2}+B^{-}dz^{2}
\end{equation}
Once this fact is established, we can repeat the same reasonings as in the
case of charged nonrotating black holes, and arrive at the same conclusions.
Namely, on the horizon (say, $U=0$) a usual particle has $V\neq 0$ whereas
the critical one has $V=0$ (that corresponds to the bifurcation point), so
the collision does not occur. For other pairs of particles (two critical or
two usual ones) collision can occur but with the finite energy $E_{c.m.}$.
This generalizes observations made for the particular case of the Kerr
metric in \cite{lake2} and \cite{gp}. This means that the strong version of
the BSW effect is not possible. However, the weak version is admissible. The
same conclusions apply to the PS effect.

\section{How nature escapes infinite energies: strong version of effect and
kinematic censorship}

Obviously, in any physical process an infinite energy cannot be released, so
some mechanism should act that prevents the strong version of the effect and
excludes the events in which such an energy could be otherwise realized. For
collisions on the event horizon, such mechanism consists in impossibility to
reach the extremal horizon within a finite interval of the proper time. In
the nonextremal case, the proper time is always finite but the critical
particle cannot reach the horizon because of the potential barrier. A\
near-critical particle can do it but the corresponding energy $E_{c.m.}$ is
finite although can become larger and larger as the state of a particle
comes closer and closer to the critical one. (See \cite{gp}, \cite{prd}, 
\cite{gpgc} for details.).

For the inner horizon, neither of two aforementioned mechanisms which were
valid for the event horizon, now applies. The inner horizon is nonextremal
by its very meaning, so the proper time needed to reach it, is finite. Apart
from this, there is no potential barrier between a near-critical particle
and the horizon. Instead, now quite different reason makes collision with an
infinite energy $E_{c.m.}$ impossible. As far as the BSW effect is
concerned, an infinite energy $E_{c.m.}$ requires that one of two particle
be critical while the other one must be usual.\ Then, it turns out that such
particles cannot meet each other in the same point of space-time. This is
because the first particle passes through the bifurcation point where
horizons intersect whereas the other one does not. This means that some kind
of censorship (let us call it "kinematic censorship") indeed acts in these
processes forbidding infinite energies in any collision but its
manifestation is quite different for the event horizons and the inner ones.
It is worth noting that the claim of \cite{ch2} are in contradiction with
kinematic censorship. As regards, the PS effect, the requirement of an
infinite energy $E_{c.m.}$ enforces colliding particles to hit the different
branches of the horizon. Thus both in the BSW and PS effects the particles
turned out to be separated in space-time although the details of such a
separation are different.

\section{Conclusion}

The situation with collisions on the inner horizon proved to be more diverse
than simply impossibility of the BSW effect. It required careful distinction
between two different effects (the BSW and PS ones) and examining two
versions of each of them (strong and weak). We checked that the strong
version is impossible and interpreted it as a particular realization of
"kinematic censorship" which has different manifestations on the event and
inner horizons. The weak version of both effects can be realized on the
inner horizon. It is worth paying attention that in the BSW effect one of
particles should be critical or near-critical while the other one should be
usual. In this respect, the situation is quite similar to what happens for
the high-energy collisions near the event horizon. Meanwhile, geometrically,
the critical and usual particles change their mutual role with respect to
the local light cone as compared to the BSW effect near the event horizon.
There is one more difference. For the BSW effect to occur near the
nonextremal event horizon, multiple scattering is required in the course of
which a particle changes its angular momentum, overcomes the potential
barrier and gets a critical value in the vicinity of the horizon. Meanwhile,
the inner horizon is not surrounded by a potential barrier, so multiple
scattering is not necessary for the BSW effect to occur there.

Also, we showed that inside the event horizon the PS effect becomes as
physically relevant as the BSW one. For the PS effect, fine-tuning the
parameters of one particle to the critical values is not necessary.

The results obtained in the present paper are valid not only for the Kerr
metric but also for generic spherically symmetric black holes as well as
generic rotating ones.

It is known that there are instabilities inherent to inner horizons (see,
e.g., the reviews \cite{ori}, \cite{and} and references therein). One can
ask whether ultra-high energy collisions (even with finite but growing
energy) expand the list of these instabilities.

\section{Acknowledgement}

I thank Yuri Pavlov for interest to this work and stimulating discussion.

\newpage \bigskip Figure 1. Impossibility of the strong version of the BSW
effect. The critical particle 1 passes through the bifurcation point whereas
a usual one 2 hits the left horizon.

\bigskip Figure 2. The weak version of the BSW effect. Near-horizon
collision between critical particle 1 and usual one 2.

\bigskip Figure 3. Impossibility of the strong version of the PS effect. The
critical particle 1 passes through the bifurcation point whereas a usual one
2 hits the left horizon.

\bigskip Figure 4. Impossibility of the strong version of the PS effect. Two
usual particles hit different branches of the horizon.

\bigskip Figure 5. The weak version of the PS effect. Near-horizon collision
between critical particle 1 and usual one 2.

\bigskip Figure 6. The weak version of the PS effect. Collision between two
usual particles near the left horizon.

\bigskip Figure 7. The weak version of the PS effect. Collision between two
usual particles near the bifurcation point.

\newpage

\begin{figure}[tbp]
Fig.1
\par
\begin{center}
\includegraphics[height=3in, bb=59 242 456 707]{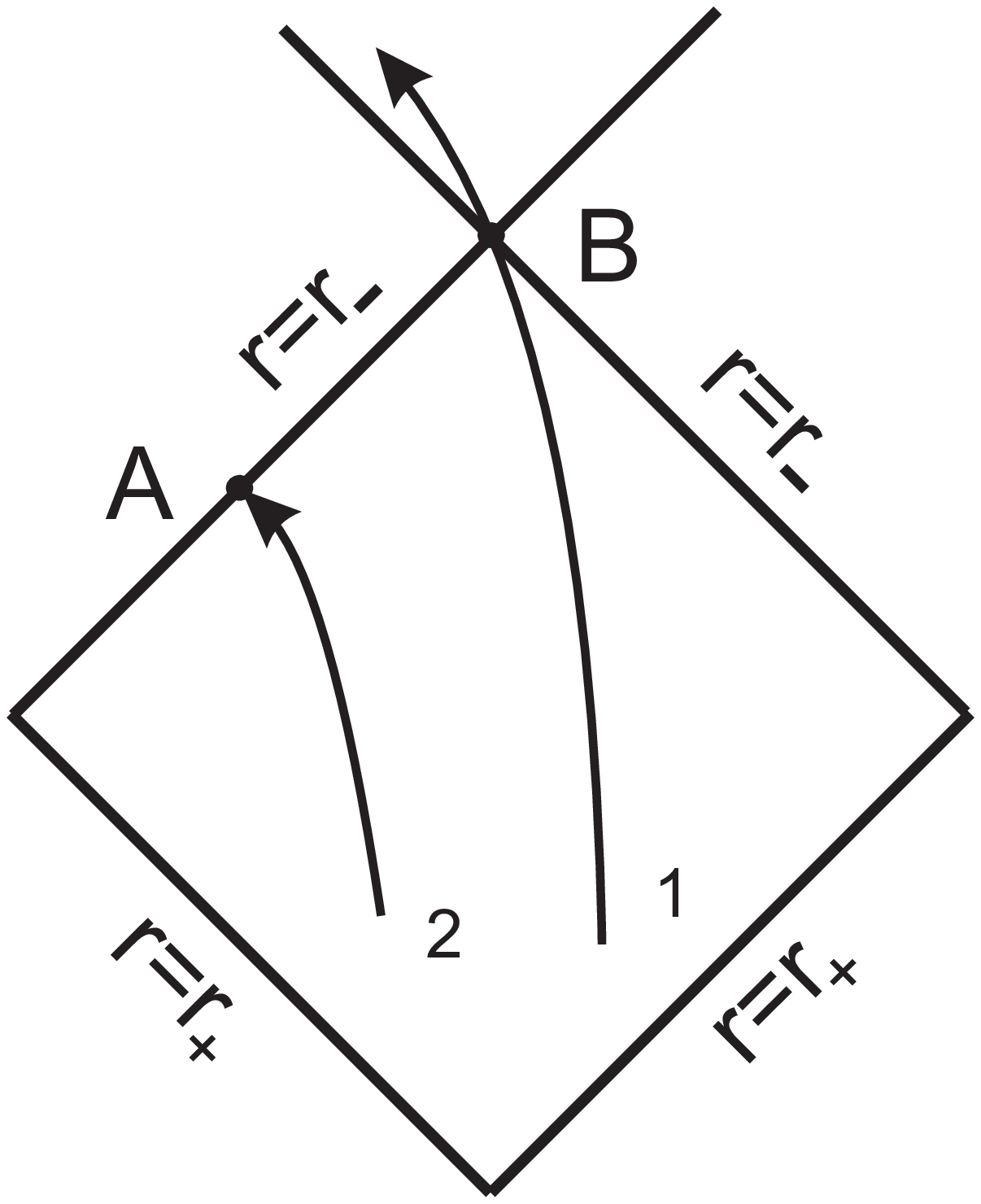}
\end{center}
\end{figure}

\begin{figure}[tbp]
Fig.2
\par
\begin{center}
\includegraphics[height=3in, bb=59 242 456 707]{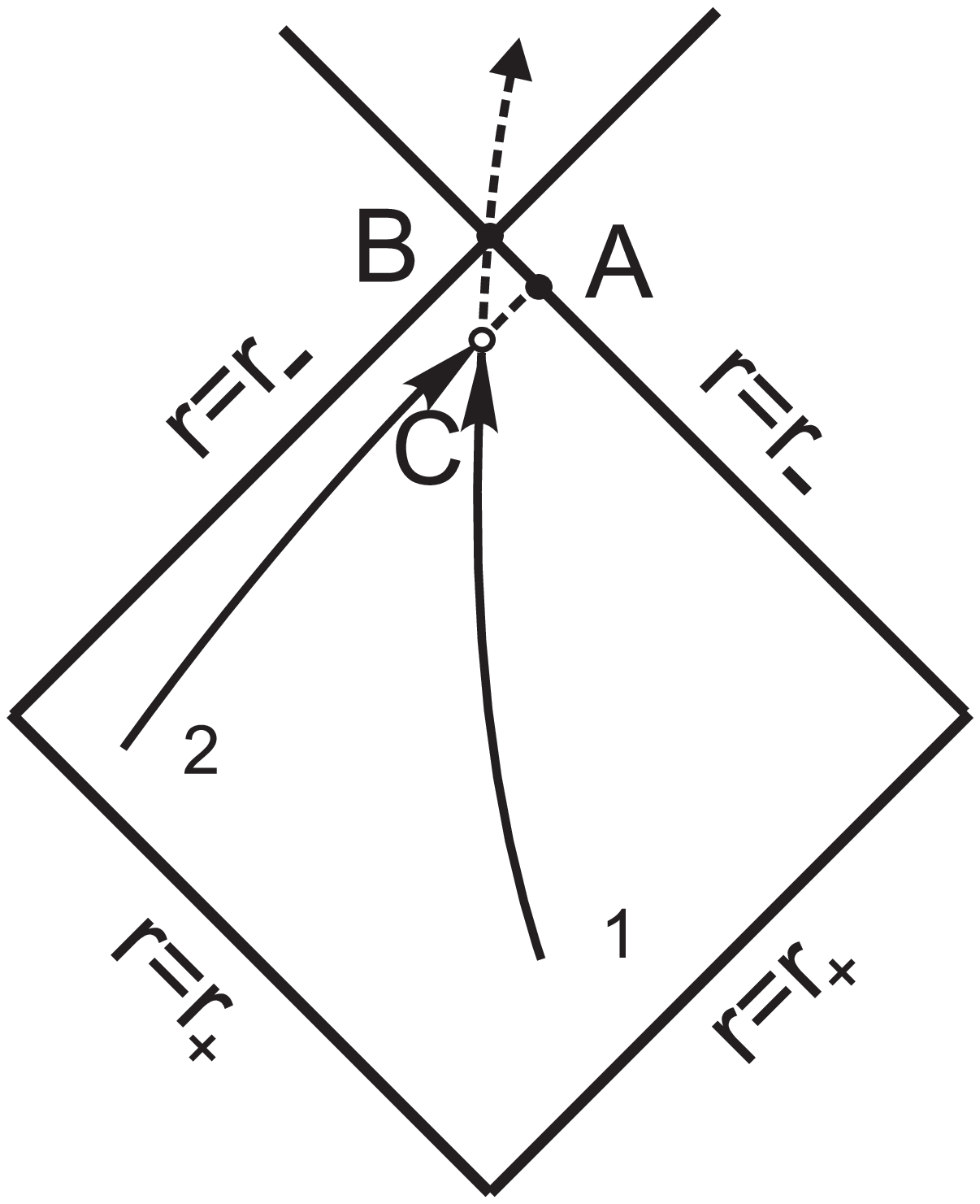}
\end{center}
\end{figure}

\begin{figure}[tbp]
Fig.3
\par
\begin{center}
\includegraphics[height=3in, bb=59 242 456 707]{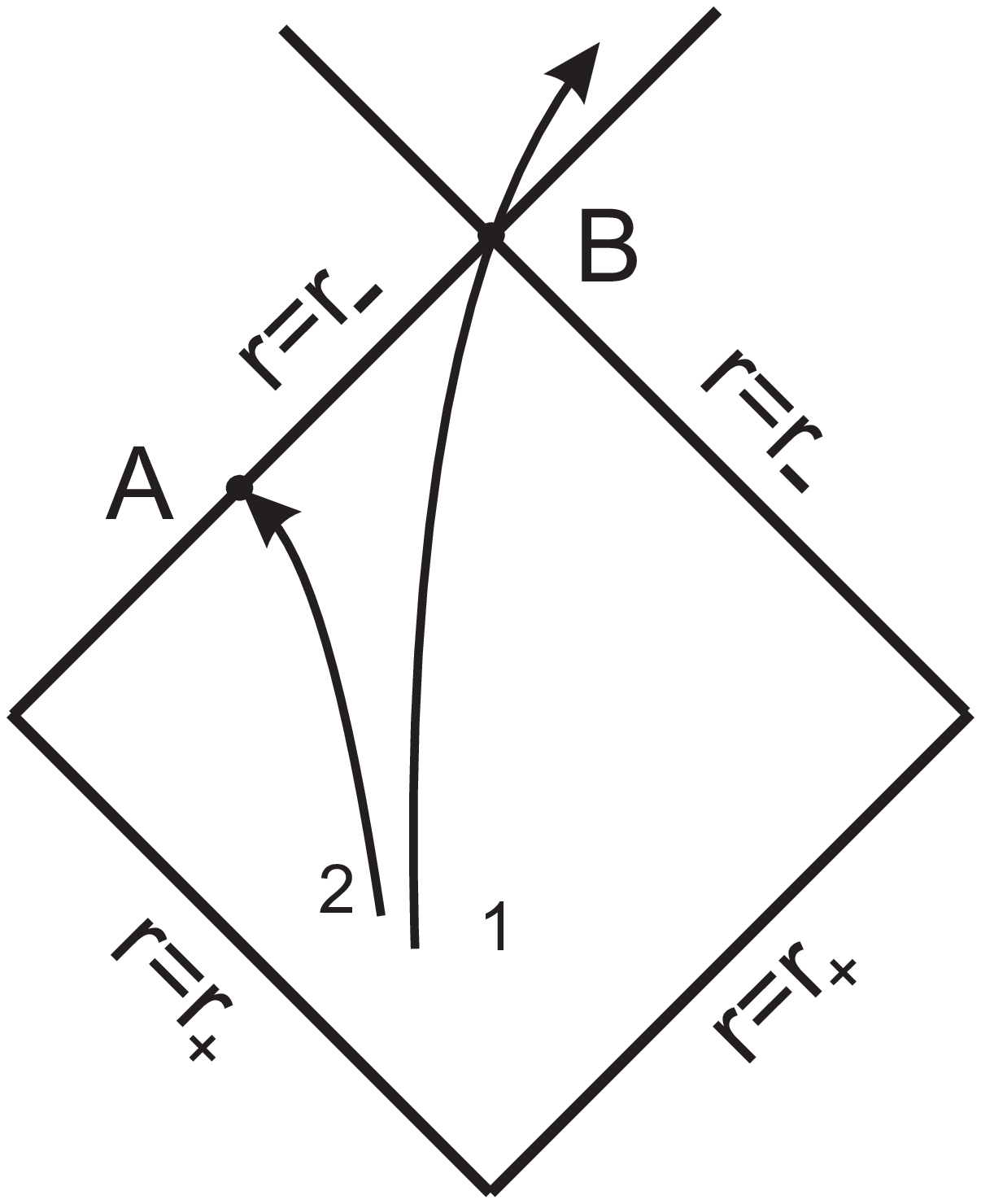}
\end{center}
\end{figure}

\begin{figure}[tbp]
Fig.4
\par
\begin{center}
\includegraphics[height=3in, bb=59 242 456 707]{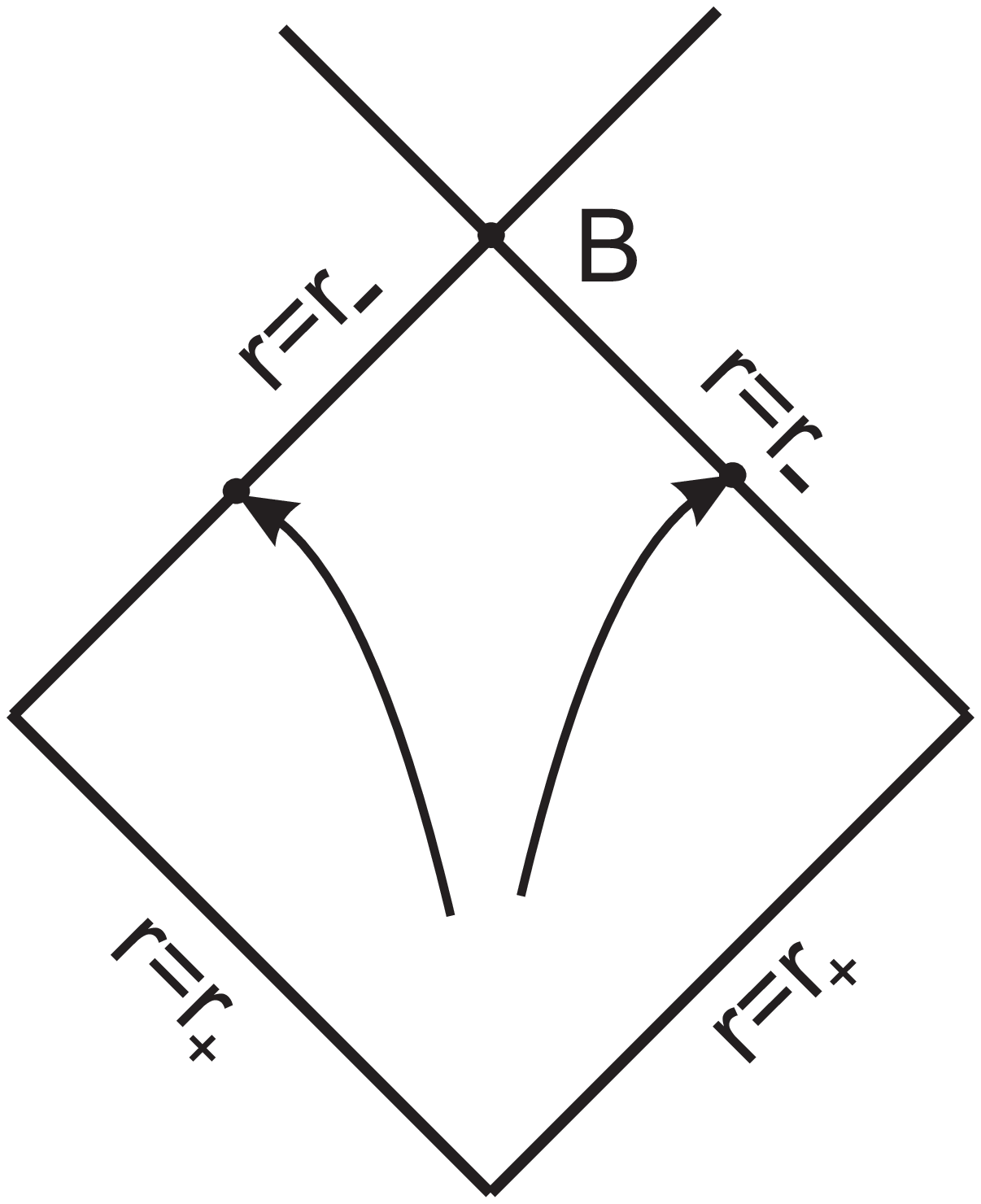}
\end{center}
\end{figure}

\begin{figure}[tbp]
Fig.5
\par
\begin{center}
\includegraphics[height=3in, bb=59 242 456 707]{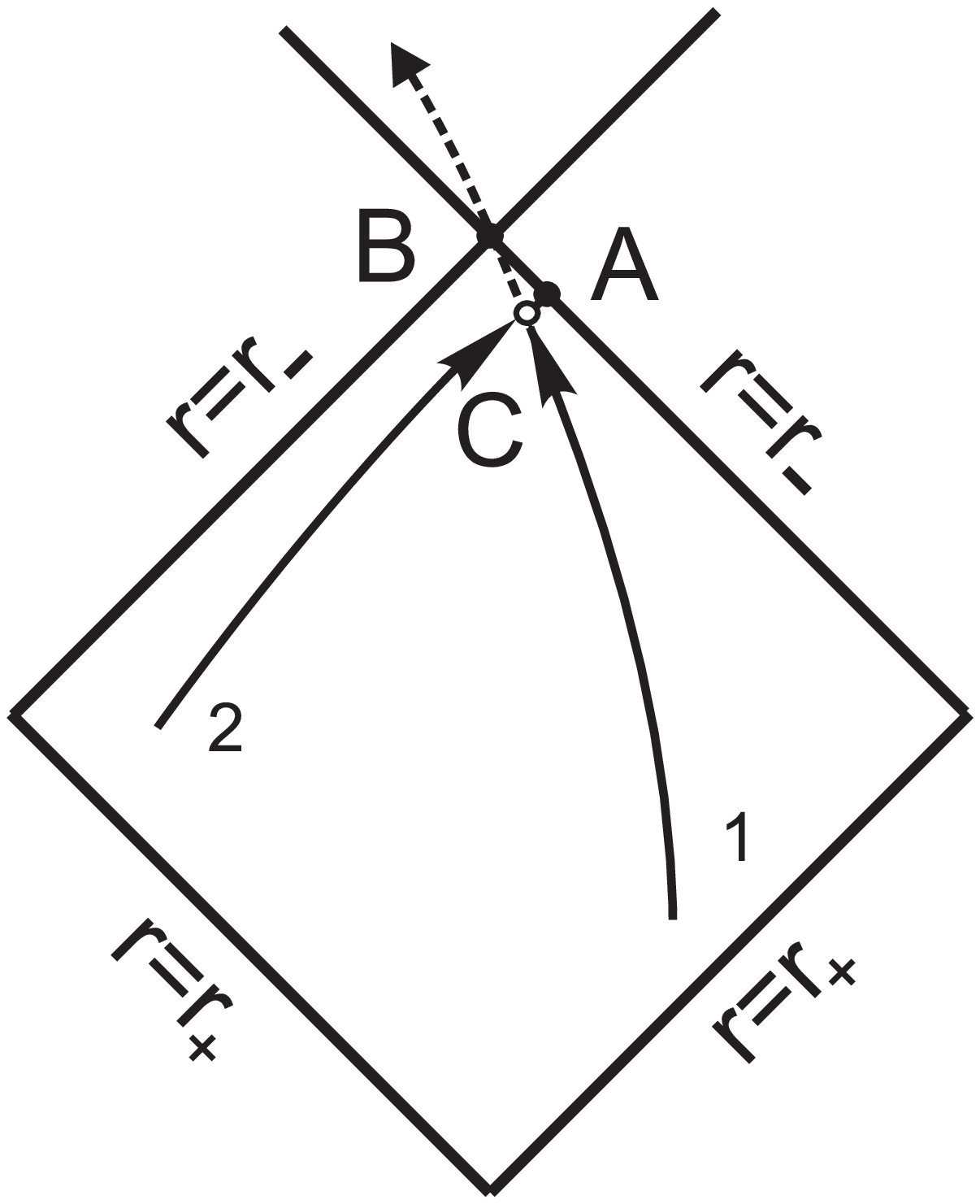}
\end{center}
\end{figure}

\begin{figure}[tbp]
Fig.6
\par
\begin{center}
\includegraphics[height=3in, bb=59 242 456 707]{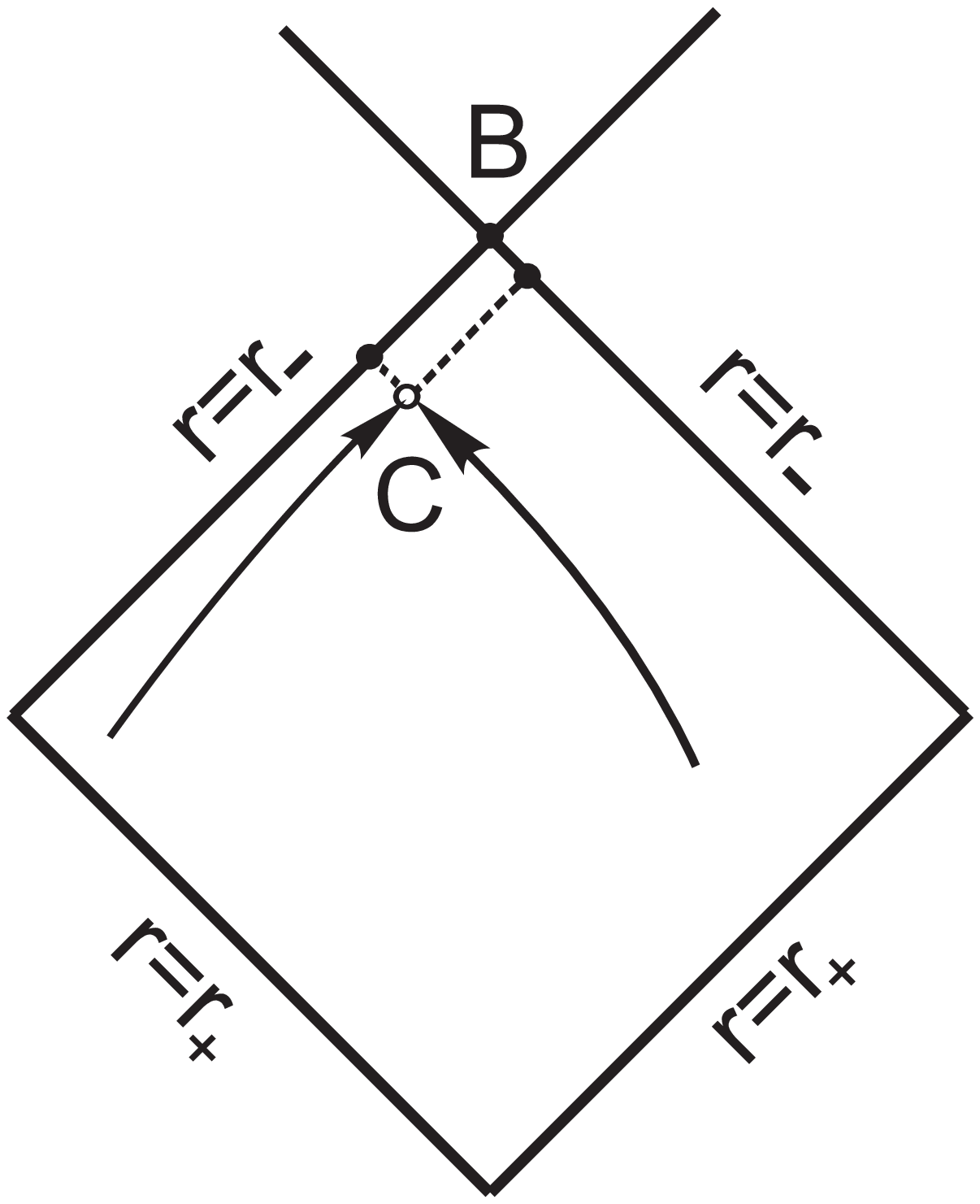}
\end{center}
\end{figure}

\begin{figure}[tbp]
Fig.7
\par
\begin{center}
\includegraphics[height=3in, bb=59 242 456 707]{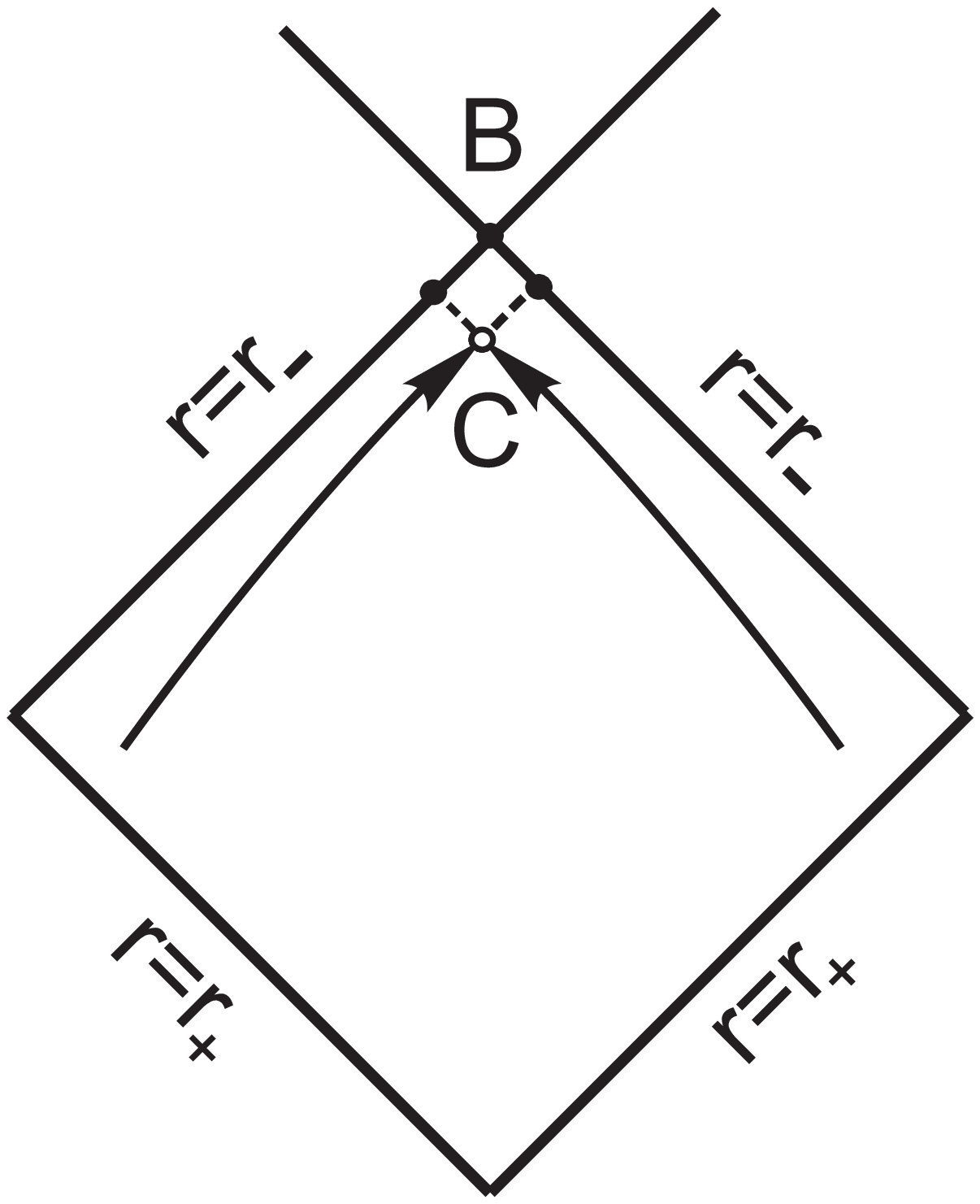}
\end{center}
\end{figure}

\end{document}